# EFFECT OF ELECTRON SCREENING ON THE α DECAY

F. F. Karpeshin, M. B. Trzhaskovskaya, L. F. Vitushkin

D.I.Mendeleyev Institute for Metrology

## Introduction

There is no need to say that these effects have great consequences for astrophysics, as well as for experiments with plasma and others in the Earth conditions. The α decay plays a significant role both in practical applications and scientific research. At the same time, there is a contradiction between the laboratory study, which considers nuclear reactions with no respect to a possible role of the electron shell or environment, and other applications, which deal with various electronic shells or environments. Specifically, the alpha decay affects the nuclear synthesis in the stars.

Surprisingly, up to paper [1] of 2013, everybody considered this question within the framework of the frozen electron shell (FS) model. This was against simple arguments, which can be readily told out based on the physical ground. The resultant effect obtained appears to be even of the wrong sign. Electron environment *retards* the decay. This might promote formation of heavy elements and actinides in the *r*-process.

Within the framework of the Gamow theory, a conventional expression for the *α* decay probability is essentially given by the product of two factors: the cluster preformation and the penetration probabilities. The former is not affected by the shell. The second factor *P* is determined by action *S*. For a bare nucleus it reads as follows:

$P = e^{-2S}$, with

$$S = \int_{R_1}^{R_2} \sqrt{2m[E - V(r)]} \, dr , \qquad (1)$$

$R_1$, $R_2$ being the turning points, *E* – the energy of the *α* particle, *V(r)* – the *α*-nucleus potential, including the strong, Coulomb and centrifugal interactions. The problem is therefore to involve the interaction of the α particle with the electronic shell and to take into account a decrease of the *Q* value *ΔQ* ≈ 40 keV in the energy of the α particle at the asymptotic region. The latter leads to the hindering of decay, in contrast to what could be expected in view of the attraction between an α particle and electrons, though this has no effect on the dynamics of alpha decay as such. At the same time, the problem of consistently taking into account the aforementioned factors became a stumbling block for many early calculations. The study of Patyk and his coauthors [2], who performed the most consistent calculations of alpha decay for the example of the chain of radon atoms, put a period to the development of the FS model.

## Physical premises

Passing the electron shell, the α particle performs a work *W*, which goes for rearrangement of the shell. According to the Feynman-Hellman theorem, *W* can be expressed as follows:

$$W = \int_{R_i}^{\infty} -\frac{\partial U^{(ad)}(r)}{\partial r} dr = -U^{(ad)}(R_i) = \Delta B ,$$

where *ΔB* is the difference of the binding energies of the initial and final atoms. This work is subtracted from the kinetic energy of the *α* particle. As a result, there is no gain in the neutral atom, as the negative potential at the starting point $R_i$ on the nuclear surface becomes strictly compensated by the decrease of the kinetic energy in (1). It is the gradient of the potential energy which is important on the remaining part of the trajectory, as the force acting on the particle is $F = -\nabla U^{(ad)}(r)$, and the force is retarding. It thus decreases the penetration probability through the barrier and the decay rate. Furthermore, taking into account the equality $\Delta Q = \Delta B$, one can substitute into (1) $U^{(ad)}(r)$ with subtracted value at the starting point $R_i$, simultaneously holding the $Q$ value. Thus, such a reduced electronic potential appears merely as the extra potential, which affects the barrier penetration probability during the $\alpha$ passage through it. In what follows, we will mean precisely the potential thus reduced under alpha-electron interaction. The resulting expression for the modified action becomes

$$S = \int_{R_1'}^{R_2'} \sqrt{2m[E - V(r) - U^{(ad)}(r)]} \, dr ,$$

with the turning points modified in accordance with adding the reduced potential $U^{(ad)}(r)$. This makes further physical consideration ultimately clear and transparent [1,3,4].

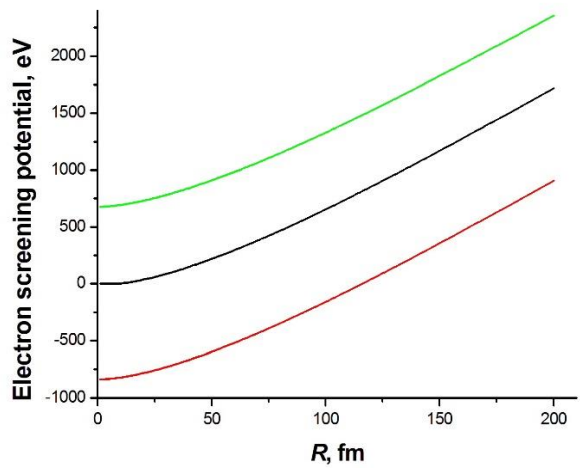

In Fig. 2 the resulting potential acting on the $\alpha$ particle from the electron shell is presented in more detail. The adiabatic potential (black curve) is compared with the potentials of the frozen shell in the initial (red curve) and final (green curve) atoms, respectively. The latter potentials are calculated by means of the RAINE package of the computer codes [5]. The adiabatic potential takes into account the rearrangement of the shell occurring due to displacement of the $\alpha$ particle. Evidently, it is between the frozen potentials of the initial and final atoms. The rearrangement energy is added to the potentials, so that the curve corresponding to the adiabatic potential starts from the zero value. Then the frozen potentials start from the negative and positive values in the initial and final atoms, respectively. Note, that the Feynman-Hellman theorem does not apply to the FS model. The left and right barrier turning points are $R_1$ = 9 fm, $R_2$ = 51 fm. The three conclusions drawn below clearly follow the fig. 2.

1) The effect of the shell on the α decay rate is negative (the adiabatic curve).
2) Contrary, within the framework of the FS model, the effect is positive (the lowest curve).
3) In the subbarrier region, the absolute value of the adiabatic potential is considerably less than that of the FS model. This explains why the effect within the consecutive adiabatic approach turns out to be not only of opposite sign, but also by an order of magnitude less.

**Results**

Calculations of the effect of the electron screening on the α decay rate were performed in the adiabatic approximation in [1,3,4]. Some of the results are presented in the Table.

**Table.** Calculated relative change $Y$ in the alpha-decay rate for bare nuclei (last column)

| Nucleus | $Q$, MeV | $T_{1/2}$ | $Y$, % |
|---|---|---|---|
| $^{144}_{60}$Nd | 1.905 | $2.29 \times 10^{15}$ yr | 0.24 |
| $^{214}_{86}$Rn | 9.208 | 0.27 $\mu$s | 0.02 |
| $^{226}_{88}$Ra | 4.871 | 1600 yr | 0.23 |
| $^{252}_{98}$Cf | 6.217 | 2.645 yr | 0.28 |
| $^{241}_{99}$Es | 8.320 | 9 s | 0.12 |
| $^{294}$118 | 11.65 | 0.89 ms | 0.27 |

Comparison of the values calculated for these nuclei with one another shows that the effect of screening strongly decreases with increasing $Q$ value and, correspondingly, decreasing $T_{1/2}$. Within the adiabatic approximation, it is evident that the most inner electrons produce more effect, as they are more sensitive to the motion of the α particle in the subbarrier area near the nucleus. According to [3], more than 80% of the effect are produced by the $K$ electrons. This suggests an elegant and basic way of experimental check e.g. through measurement of the difference in the decay rate between the He-like and bare ions of the same nuclei-alpha emitters. Monochromaticity parameters of the storage ring beam are good enough, in order to detect the recoil of the nuclei as a result of alpha decay e.g. by the Schottky method. A list of candidates can be proposed as follows:

- $^{212}$Rn: 23.9 m (100% α);
- $^{212}$Fr: 20 m (43% α);
- $^{212}$Ra: 13 s (90% α);
- $^{211}$Ra: 13 s (93% α).

Such lifetimes for seconds to minutes should be most convenient and quite suitable for experiment. The latter can be realized with similar technics which was applied to conclude about time modulation in beta decay of Pm ions [6]. The daughter product from the alpha decay will likely stay in the ring and should be seen by Schottky analysis. If one starts with $^{212}$Rn$^{84+}$, one should have most of the time $^{208}$Po$^{82+}$, it will be good to detect both. If one uses bunches of ions then the Schottky needs to be calibrated and one has to find out how well this can be done for the four cases of decay from $^{212}$Rn$^{84+}$ and $^{212}$Rn$^{86+}$. Concerning the "background" by ionization and shake-off from Rn$^{84+}$, and there could also be electron capture by Rn$^{86+}$, these could be made quite weak.

**Conclusion**

Although the predicted magnitude of the effect is as small as about $10^{-3}$, there is a circumstance favorable for experimental studies: about 80% of the effect is due to $K$-shell electrons. In order to observe the effect, it is therefore sufficient to compare the decay half-lives for bare nuclei with the half-lives for respective one and (or) two-electron atoms without invoking neutral atoms – their accumulation in accelerator rings is impossible, and it is more reliable to observe effects as small as that in question via difference measurements at the same facility. Thus, comparative measurements can be performed in the same storage rings with bare nuclei and respective ions of different multiplicity – for example, helium ions.

An experiment on comparison of the alpha decay periods in neutral atoms and bare nuclei was already tested at GSI a few years ago [7]. As compared to that experiment, later papers [1,3,4] contain qualitative details making such an experiment more feasible.

First, the main idea of that experiment sounds striking: comparison of the periods of neutral atoms with those in H-like ions. Why not just bare nuclei? One *1s* electron already produces close to half the effect. Why was it needed?

Second, they compared the periods in the H-like ions with those in neutral atoms, as measured by traditional techniques in stoppers. Our proposal assumes that both periods – of the bare nuclei and He-like ions – will be measured in the same channel of the storage ring by the Schottky method. This way will considerably compensate systematic uncertainties, which will be merely the same, whereas in the previous experiment they could be of different origin and therefore, not correlated.

Third, a general idea was told out in [7] that such an experiment is feasible with the resulting uncertainty within 0.1% [2], which one can think opens green light to its realization.

The authors are grateful to Yu. Litvinov, X. Ma, T. Stölker and R. Schuch for fruitful discussions.